# Interplay of Stimulated Emission and Fluorescence Resonance Energy Transfer in Electrospun Light-Emitting Fibers

*Lech Sznitko,[†] Luigi Romano,[§,††] Andrea Camposeo,[††] Dominika Wawrzynczyk,[†] Konrad Cyprych,[†] Jaroslaw Mysliwiec,[†] and Dario Pisignano[§,††,¶,*]*

[†] Wroclaw University of Science and Technology, Wybrzeze Wyspianskiego 27, 50-370 Wroclaw, Poland

[§] Dipartimento di Matematica e Fisica "Ennio De Giorgi", Università del Salento, via Arnesano I-73100, Lecce, Italy

[††] NEST, Istituto Nanoscienze-CNR, Piazza S. Silvestro 12, I-56127 Pisa, Italy

[¶] Dipartimento di Fisica, Università di Pisa, Largo B. Pontecorvo 3, I-56127 Pisa, Italy

[*] Corresponding author

**ABSTRACT**

Concomitant amplified spontaneous emission (ASE) and Förster resonance energy transfer (FRET) are investigated in electrospun light-emitting fibers. Upon dye-doping with a proper FRET couple system, free-standing fibrous mats exhibit tunable FRET efficiency and, more importantly, tailorable threshold conditions for stimulated emission. In addition, effective scattering of light is found in the fibrous material by measuring the transport mean free path of photons by coherent backscattering experiments. The interplay of ASE and FRET leads to high control in designing optical properties from electrospun fibers, including the occurrence of simultaneous stimulated emission from both donor and acceptor components. All tunable-optical properties are highly interesting in view of applying electrospun light-emitting materials in lightening, display, and sensing technologies.

## 1. Introduction

The search for efficient and cheap light-emitting and lasing materials based on macromolecular materials has been increasingly interesting for scientists and material engineers. In the last decade, organic and polymeric systems have fully shown their potential for building lasers,[1-5] light-emitting devices[6-8] and transistors,[9-11] optical sensors,[12,13] and electrochromic components.[14-16] Many of these applications critically benefit from miniaturizing active organic structures, which can lead to a variety of effects including preferential orientation of macromolecules along the longitudinal axis of nanofibers,[17-19] optical anisotropy,[20] increased charge-transport,[21,22] and enhanced mechanical properties.[23] For instance, the generation of organic nanowires and nanofibers through electrospinning involves the anisotropic stretching of a polymer solution, assisted by an external electric field.[24-26] In this way, fibers can be realized, with diameters ranging from a few microns down to a few nm. The technique is simple and cheap, and versatile in terms of usable polymers and solvents. Thanks to the anisotropic geometry of polymer nanofibers, photons generated by light-emitting materials can be waveguided[27] and a light signal can be amplified along the polymer structure.[28]

Organic nanofibers are also been considered as active materials in new laser configurations, such as so-called random lasers, in which the feedback for device operation is delivered by multiple scattering of light instead of two mirrors as in classic resonators.[29,30] These devices can operate in two regimes, the first one very similar to classical lasers is called coherent regime. This type of lasing is supported when scattering is strong enough to establish light localization and the photon diffusion is partially or completely stopped by interference effects. The second type is called incoherent random lasing, which occurs when multiple scattering is responsible for light diffusion. In such case photons are going to follow Brownian motion, possibly experiencing higher gain compared to ballistic transport due to the higher residence time in the active material. For these reasons, assessing the strength of scattering in light-emitting molecular nanomaterials, which also show optical gain, is highly interesting in view of discriminating different lasing regimes and better addressing applications, which might include lightening, imaging and microscopy. For instance, it has been recently shown that the quality of microscope imaging might be improved by the use of incoherent random lasers or amplified spontaneous emission (ASE) sources, which do not lead to speckle patterns related to interference.[31]

In this paper we study for the first time the concomitant ASE and Förster resonance energy transfer (FRET) in electrospun light-emitting fibers through experiments carried out on a series of free-standing mats doped with two different laser dyes, i.e. rhodamine 6G (Rh6G) and cresyl violet (CrV). The obtained fibers show tunable FRET efficiency and, more importantly, tailorable optical threshold conditions for stimulated emission, as well as effective scattering of light with features characteristics of incoherent random lasing as assessed by coherent backscattering (CBS) measurements. The interplay of ASE and FRET leads to the occurrence of simultaneous stimulated emission from both the donor and the acceptor components under properly designed conditions. For their set of unusual optical properties, all tunable by controlling the electrospinning process parameters, this class of electrospun fibers might be highly useful for the development of stable light-emitting and lasing organic systems, with application in display and sensing technologies.

## 2. Experimental Section

*2.1. Light-emitting fibers.* Poly(methyl methacrylate) (PMMA, number-average molecular weight, $M_n$ = 120,000 g/mol), chloroform, and *N,N*-dimethylformamide (DMF) are obtained from Sigma-Aldrich. Rh6G and CrV are from Exciton. The chemical structures of the two dyes are shown in Fig. 1a and 1b, respectively. Solutions are prepared by dissolving 300 mg of PMMA in 1 mL of a mixture of chloroform and DMF (4:1 v:v). Rh6G and CrV are added to the solution keeping the total dye concentration at 1% (w:w) compared to PMMA, and by relative Rh6G/PMMA amounts of 1% (corresponding to a CrV/Rh6G, acceptor:donor relative molar concentration, $\phi = 0$), 0.87% ($\phi = 0.2$), 0.69% ($\phi = 0.6$), 0.5% ($\phi = 1.3$), 0.31% ($\phi = 2.8$), 0.13% ($\phi = 8.6$), and 0% for samples named as *S*1, *S*2, *S*3, *S*4, *S*5, *S*6, and *S*7, respectively.

To electrospin fibers, solutions are placed in a syringe with a metal needle (21 gauge), and a flow rate of 0.7 mL/h is set by a programmable syringe pump (Harvard Apparatus), applying a voltage bias of 18 kV, and using a distance of 25 cm from the syringe needle to a metal collector. The deposition of each sample is performed on microscopy slides and takes about 30 min (photographs in Fig. 1c). The morphology of fibers is inspected by an optical microscope (Olympus), working in fluorescence and transmission modes with magnification 40× (Numerical Aperture, *N.A.* = 0.65). In fluorescence mode, samples are excited by a halogen lamp through a U-MWIB3 fluorescence filter cube, delivering the excitation in the spectral range between 460 nm and 495 nm. More insight in fibers morphology is achieved by scanning electron microscopy (SEM, JSM 6610LVnx by JEOL Ltd), using an acceleration voltage of 6 kV.

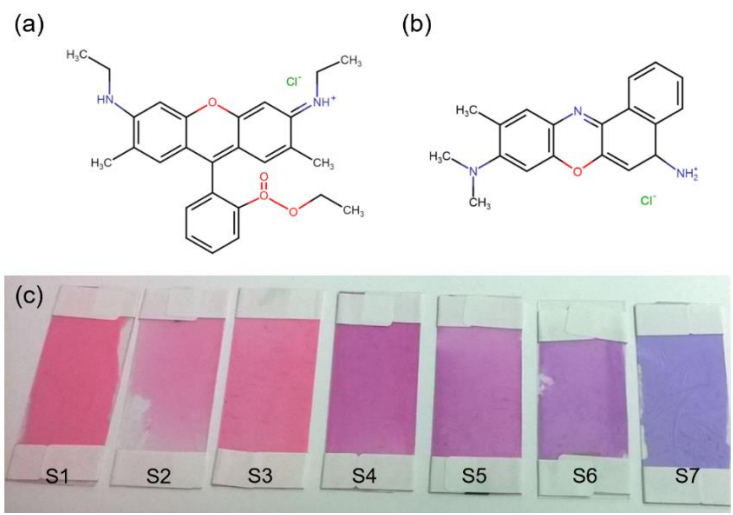

**Figure 1.** (a, b) Chemical structure of Rh6G (a) and CrV (b). (c) Series (*S1-S7*) of electrospun samples, with CrV/Rh6G ratio progressively increasing from left to right. The size of each sample is 25×75 mm$^2$.

*2.2. Spontaneous and stimulated emission.* Spectroscopic measurements are carried out using a F4500 spectrofluorometer (Hitachi), allowing us to obtain emission spectra below the excitation threshold for stimulated emission. The luminescence lifetimes are measured using the Time Correlated Single Photon Counting (TCSPC) method,[32] with the Becker&Hickl system comprising a TCSPC Module (SPC-130-EM), and a hybrid photomultiplier detector (HPM-100-06) with detector control card (DCC 100) mounted on a Princeton Instruments spectrograph (Acton SpectraPro-2300i) under excitation with picosecond, 516 nm laser diode (BDL-516-SMC). The averaged lifetime values are obtained as $\langle \tau \rangle = \frac{A_1 t_1^2 + A_2 t_2^2}{A_1 t_1 + A_2 t_2}$, where $A_1$ and $A_2$ are the exponential decay amplitudes for short and long living components, respectively, and $t_1$ and $t_2$ are the corresponding characteristic times obtained through fitting the experimental decay curves with double exponential decay functions.[33] The calculated values of emission lifetimes are then used for estimating the FRET efficiency according to the equation:[33]

$$\eta = \left(1 - \frac{\langle \tau_{DA} \rangle}{\langle \tau_D \rangle}\right) 100\%, \tag{1}$$

where $\langle \tau_{DA} \rangle$ is mean lifetime of donor emission in the presence of acceptor, and $\langle \tau_D \rangle$ is mean lifetime of donor emission without acceptor. The stimulated emission properties are assessed using an Ocean Optics USB 2000 fiber spectrometer, collecting the signal from the edge of fibers excited by a nanosecond pulsed Nd:YAG laser (Surelite II, Continuum) at wavelength ($\lambda$) of 532 nm with repetition rate 10 Hz. The power of the pumping beam is controlled by rotating the half-wave plate inserted before a Glan-Laser polarizer, with vertical transmission direction of electric field vector.

*2.3. Light-scattering.* The set-up used for light-scattering measurements is schematized in Fig. S1 in the Supporting Information. CBS experiments are performed by a cw He-Ne laser (HNS-20P-633, Meredith Instruments, $\lambda = 632.8$ nm) with vertical polarization and 20 mW power. The laser beam passes through a beam expanding system, followed by a beam splitter. The first reflected component is extinguished by the beam dump. The transmitted component with vertical linear polarization passes through the quarter-wave plate set at the azimuth of 45° with respect to polarization of incident beam. The quarter-wave plate converts linear polarization state of incident beam into right circularly polarized light. The light of such polarization state is then coherently backscattered and the light once again passes through the quarter-wave plate. In this case the right circular polarization state becomes converted into linear with azimuth oriented horizontally (perpendicular to initial state). Such beam is then reflected by the beam splitter to a motorized detection system. In order to filtrate specular reflection coming from He-Ne laser generated on beam splitter, the polarizer with horizontally oriented transmission direction is placed after the beam splitter. Finally, the light is focused by the convex lens on a 100 μm pinhole placed before a PMM02 photomultiplier (Thorlabs) mounted on an automatic moving stage. The transverse movement allows to scan the distribution of backscattered light intensity as a function of the observation angle which, can be then recalculated into mean transport free-path $L_t$:[34]

$$L_t = \frac{\lambda}{3\pi\Delta\omega},$$ (2)

where $\omega$ stands for the detection angle and $\Delta\omega$ is the full width at half maximum of the backscattered signal.

## 3. Results and Discussion

*3.1. Fiber morphology and optical properties.* It has been recently shown that Rh6G and Nile Blue (NB) can create an efficient FRET pair.[35-37] Unfortunately, NB cannot be easily dissolved in organic solvents like chloroform, thus it cannot be easily dispersed in transparent polymer matrices such as PMMA. The CrV is a chemical modification of NB that is soluble in organic solvents and also exhibits optical features suitable for utilization as FRET acceptor together with Rh6G. The morphology of CrV/Rh6G-doped electrospun fibrous non-wovens is shown in Fig. 2a and b, which show images obtained in transmission mode for samples containing 0.13% and 0.87% of Rh6G, respectively. Fig. 2c and d show the corresponding fluorescence images. It is clearly visible that the color of emission changes with different contents of donor and acceptor dyes. Since all of the samples are fabricated using the same preparation route, the morphology of obtained mats remains almost unchanged upon varying the relative dye doping level. In each case the fibers show no directional ordering and their thickness is mostly around few microns (2-5 μm) with a few filaments reaching a diameter around 10 μm.

Two exemplary SEM images of fibers containing 0.5% and 0.87% of Rh6G are presented in Fig.

2e and f, respectively, showing ribbon-shaped fibers of transversal size around 2-3 μm (long axis)

and 1 μm (short axis). An exemplary histogram showing the size distributions corresponding to the ribbon shape of the obtained fibers is displayed in Fig. S2. Bigger fibers can be occasionally found, as shown in Fig. 2f. The aspect ratio of the long to the short transversal size varies in range of 2.5 to 4 depending on the overall size of fiber.

The emission spectra are obtained by exciting the samples with the wavelength of 480 nm

aiming at addressing the Rh6G excitation band, except for the sample containing only the CrV

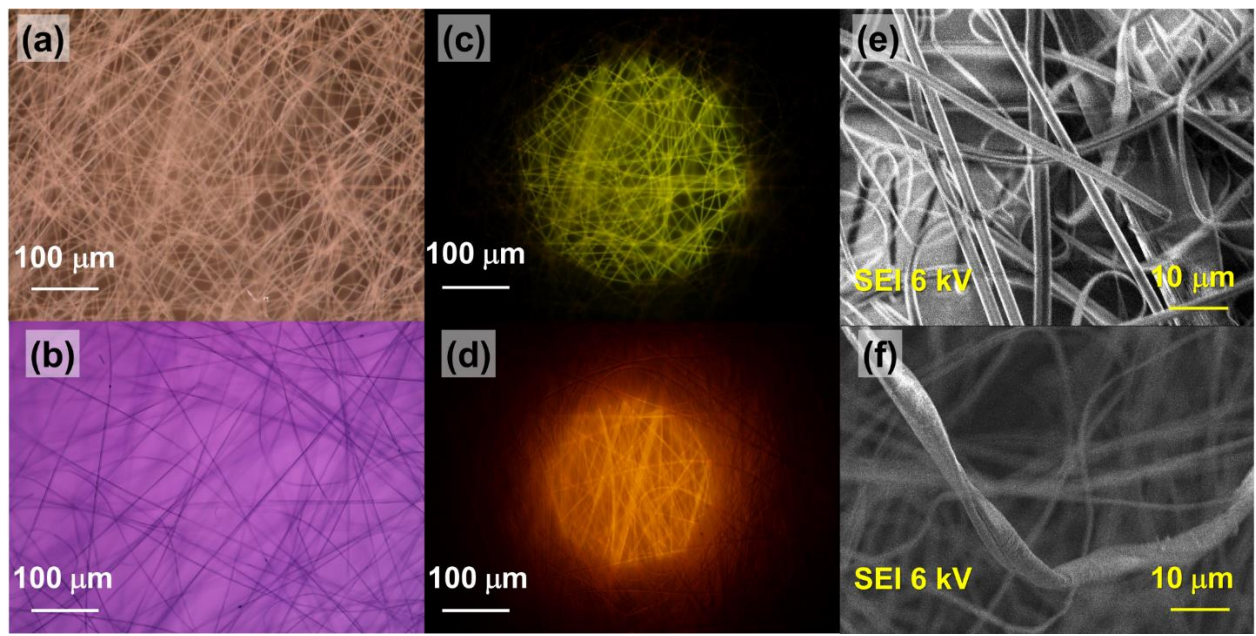

**Figure 2.** (a-d) Micrographs obtained in transmission (a,b) and luminescence (c,d) for electrospun fibers with Rh6G content of 0.13% (a,c), and 0.87% (b,d). (e,f) SEM micrographs presenting the typical morphology and the ribbon shape of obtained fibers. The Rh6G content of the samples is 0.5 and 0.87 for (e) and (f), respectively.

laser dye which is pumped at the wavelength of 560 nm. The excitation spectra are obtained by monitoring the emission intensity collected at the wavelength 660 nm, to ensure that only the CrV emission is measured, and by changing the excitation wavelength within the range between 420 nm and 630 nm. Fig. S3 shows the excitation and emission spectra for fibers doped only with Rh6G (1%) or CrV (1%). The emission of CrV is peaked at 623 nm, with a broad feature reaching the neat infrared region above 700 nm. The excitation is also broad starting form nearly 450 nm and reaching the maximum at 604 nm, thus effectively overlapping with the emission of Rh6G (540-650 nm) whose maximum is found at 573 nm. Finally, the excitation spectrum of Rh6G has its maximum at a wavelength of 537 nm, and it is spreading between 450 nm to nearly 600 nm. The two dyes are therefore excellent candidates to highlight a FRET effect in the electrospun fibers.

The emission and excitation spectra measurements obtained for fibers with various relative concentrations of CrV and Rh6G laser dyes are shown in Fig. 3. In emission spectra, it is clearly visible that above a CrV content equal to 0.13% ($\phi = 0.2$), the red-side shoulder becomes better pronounced and the CrV emission band (623 nm) arises from the underlying emission of Rh6G. Interestingly, upon reducing the Rh6G concentration and increasing concentration of CrV, the emission band related to Rh6G becomes significantly blue-shifted, with the peak varying from 575 nm to 547 nm. This phenomenon can be associated to a correspondingly increasing reabsorption by CrV molecules, or due to a decreasing population of emissive aggregates including J-dimers or higher-order structures in the electrospun fibers.[38-40]

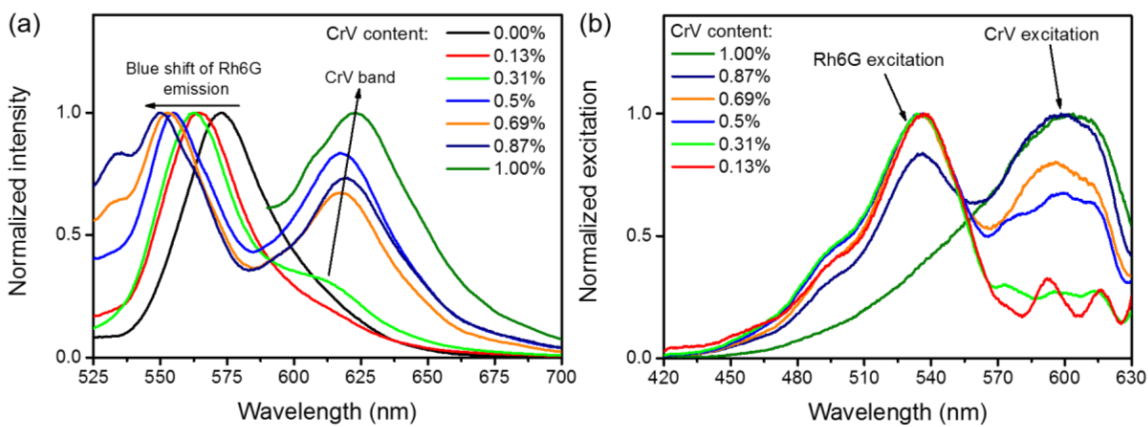

**Figure 3.** Emission (a) and excitation (b) spectra of electrospun fibers with increasing CrV and decreasing Rh6G content. Excitation wavelength in (a): 480 nm. The sample containing only the CrV laser dye is instead excited at the wavelength of 560 nm.

*3.2. FRET efficiency.* The FRET efficiency is calculated using Eq. (1), and by double-exponential fitting parameters from the emission decay curves collected at the wavelength 560 nm (Fig. 4a), highlighting the shortening of emission lifetime with increasing content of the acceptor dye (fitting parameters are presented in Table S1 in the Supporting Information). The dependence of the resulting FRET efficiency on the acceptor concentration in the fibers is shown in Fig. 4b. At the weight:weight relative concentration (donor to acceptor) 1:1 (CrV concentration 0.5%, $\phi = 1.3$), the shortening of the emission lifetime is most clearly visible, resulting in the highest FRET efficiency reaching the value of 57%. Further increase of the acceptor contents leads to a slight lowering of the FRET efficiency to the value of 49%, which might be explained in terms of depletion of donor aggregates that would help in transferring the energy to the acceptor molecule. In this respect, it is essential to note that the total concentration of the dye (CrV+Rh6G) is kept at the same level of 1%, the CrV content increases thus corresponding to a decrease of donor concentration in turn disfavoring the formation of different types of aggregates in the fiber matrix. Consequently, the shortening of the Rh6G molecular band lifetime through the interaction with aggregates is less likely at higher CrV contents. This hypothesis will be better supported in the following related to stimulated emission measurements.

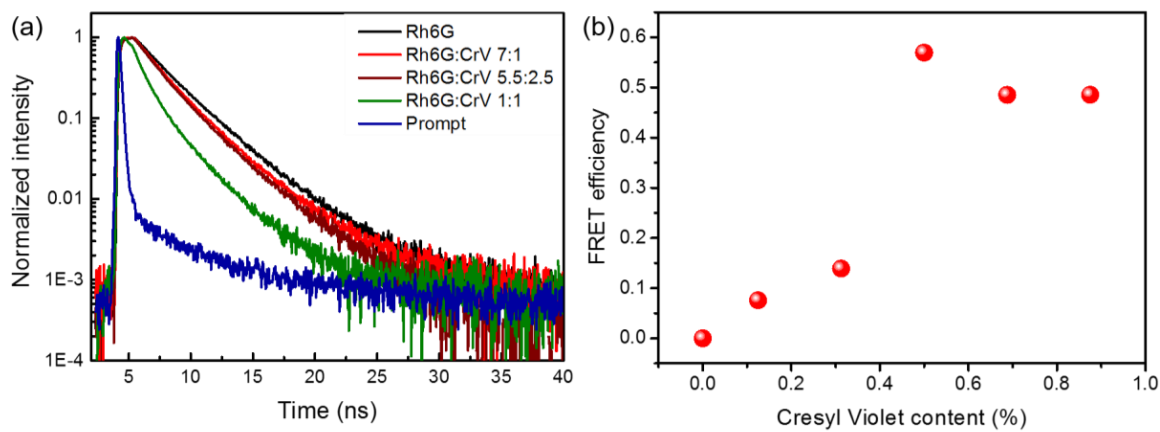

**Figure 4.** (a) Fluorescence emission (560 nm) lifetime for different Rh6G to CrV ratios. The 'prompt' profile indicates instrument response characteristics obtained for ps diode-laser excitation (b) FRET efficiency dependence on CrV relative content.

*3.3. Stimulated emission and FRET-ASE interplay.* The stimulated emission from the electrospun fibers is measured for each sample at varying CrV and Rh6G concentrations, finding different threshold levels (Fig. 5). The threshold values are estimated from plots of the emission intensity versus pumping energy, through the intersection of the two approximately linear behaviors for the low and high parts of these curves, as shown in the Supporting Information file (Figures S4 and S5). For fibers with $\phi = 0$ (no content of CrV), the stimulated emission peak is at the wavelength of 601 nm (Fig. 5a), with a threshold of 0.78 mJ/cm$^2$. The stimulated emission is then blue-shifted to 597 nm upon increasing the pumping fluence, with a simultaneous decrease of the full width at half maximum (FWHM) reaching 9 nm. When the CrV concentration is 0.13% ($\phi = 0.2$), the threshold has almost the same value equal to 0.79 mJ/cm$^2$ but the stimulated emission located at the wavelength 596 nm has a broader spectrum (FWHM = 20 nm). Moreover, the spectral shape clearly suggests the appearance of two bands underneath.

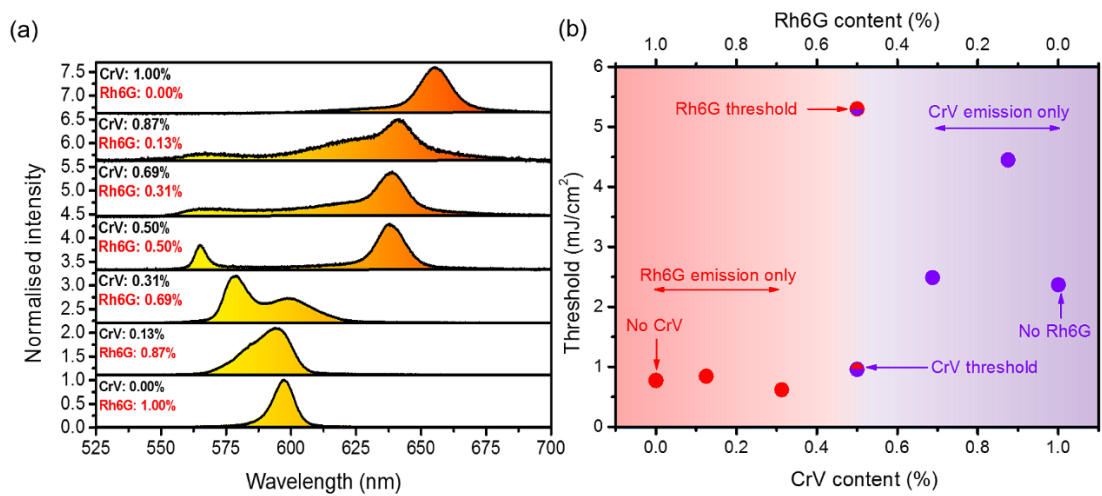

**Figure 5.** (a) Stimulated emission spectra for electrospun fibers with varying concentration of CrV (acceptor) and Rh6G (donor). (b) Threshold dependence on CrV (bottom horizontal axis) and on Rh6G (top axis) content.

When the relative concentration of CrV further increases to 0.31% ($\phi = 0.6$), the threshold for stimulated emission lowers to 0.62 mJ/cm$^2$, which is around 2 times lower than the value reported by Krämmer and co-workers for random fiber networks.[41] The shape of the stimulated emission spectrum is made more complex because two maxima are clearly emerging. The first one is located at the wavelength 578 nm, having stable position with the varying pumping fluence. The second maximum is located at the wavelength of 617 nm, when the used excitation power is near the threshold, and it changes the position to 597 nm at pumping intensities well above threshold (FWHM around 28 nm). The complexity of such Rh6G stimulated emission bands may be attributed to different aggregates involved in observed process,[40] possibly leading to a significant blue-shifted high-energy component in the stimulated emission spectra as here found by further lowering the Rh6G concentration in the light-emitting fibers down to 0.5% (Fig. 5a). In this case two stimulated emission bands are visible, well separated by a spectral distance of 72 nm. The first band (peak at 565 nm) agrees with the emission of molecular Rh6G (typical for diluted solutions)[42] and exhibits a threshold of 5.30 mJ/cm$^2$. The second one relates to the stimulated emission of CrV, appearing at the wavelength of 637 nm above 0.96 mJ/cm$^2$. A CrV content of 0.69% ($\phi = 2.8$) in the light-emitting fibers leads to stimulated emission characteristic of CrV only (637 nm, threshold at 2.49 mJ/cm$^2$). When the CrV concentration is 0.87% ($\phi = 8.6$), the stimulated emission from CrV is found to slightly red-shift (641 nm), which can be attributed to reabsorption processes. The red-shift is accompanied by increased threshold (4.45 mJ/cm$^2$). Finally, for a 1% concentration of CrV (no Rh6G in fibers), the stimulated emission is more red-shifted towards the wavelength of 655 nm, which might be associated to aggregate formation. The threshold level is at 2.37 mJ/cm$^2$. The overall behavior of the stimulated emission threshold for varying dye concentration is summarized in Fig. 5b. Since both the dyes might be excited using the 532 nm laser light, the overall photophysics is rather complex and cannot be explained in terms of simple FRET process. Instead, a tight interplay might be envisaged between FRET and depletion of excited states by stimulated emission, with aggregate species also likely involved in cascade of energy transfer as supported by the non mono-exponential character of the emission decays.

The different stimulated emission behavior of light-emitting fibers doped with varying donor and acceptor concentrations can be interpreted by means of a quantitative model which describes ASE in presence of fast non radiative energy transfer processes.[43] The FRET involving the two dyes in the fibers occurs through a resonant and near-field dipole-dipole interaction, over a characteristic distance $R_0$ (Förster radius) of a few nm, which is given by:

$$R_0 = \left( 0.5291 \frac{\kappa^2 \eta_D}{n^4 N_{AV}} \int F_H(\nu) \sigma_D(\nu) \frac{d\nu}{\nu^4} \right)^{1/6} \tag{3}$$

where $F_H$, and $\sigma_D$ are the donor emission and the acceptor molar decadic extinction coefficient, respectively, $\eta_D$ is the donor luminescence quantum yield (0.75),[44] $\kappa^2$ is an orientational factor for the dipole moments distribution, which is 2/3 for a random distribution, $n$ is the refractive index (~1.5), and $N_{AV}$ is Avogadro's number. The energy transfer rate, $K_{ET}$, would be $(R_0/R)^6/\tau_D$ for isolated molecules separated by a distance $R$, where $\tau_D$ is the radiative characteristic time of the donor. Deviations from the $R^{-6}$ dependence of $K_{ET}$ may occur due to local aggregate formation and to different geometries of the donor-acceptor interactions.[45,46] Also, values of $\kappa^2$ different from 2/3 cannot be entirely ruled out due to possible orientation effects occurring during electrospinning at molecular scale.[17,18,20,23] In the fibers, each acceptor would interact with a distribution of donors, located at distances limited by the effective molecular radius and by the fiber radius, leading to $K_{ET,fiber} = (1/\tau_D) \int (R_0/R)^6 \rho_A \phi dV$, where $\rho_A = 3/(4\pi a^3)$ is the acceptor density, $a$ is the acceptor molecule radius, and the integral is evaluated over the volume of the organic filament. The $K_{ET,fiber}$ rate is then related to the probability, $P$, of exciton transfer from the donor to the acceptor: $P = \dfrac{K_{ET,fiber}}{(1/\tau_D) + K_{ET,fiber}}$ . In addition, the ASE intensity, generated by the

optical amplification of the spontaneously emitted light with high-enough acceptor

concentrations can be approximated as:[43,47]

$$I_D = \eta_D L I_D^s \frac{\Omega}{4} \frac{[e^{g_D l (1-P)(1-\phi)-1}]^{3/2}}{[g_D l (1-P)(1-\phi) e g_D l (1-P)(1-\phi)]^{1/2}} \tag{4a}$$

$$I_A = \eta_A L I_A^s \frac{\Omega}{4} \frac{\{e^{g_A l [\phi + P(1-\phi)]-1}\}^{3/2}}{\{g_A l [\phi + P(1-\phi)] e^{g_A l [\phi + P(1-\phi)]}\}^{1/2}} \tag{4b}$$

where $I_D^s$ and $I_A^s$ are the saturation intensity for the donor and the acceptor, respectively, $g_D$ (9

cm$^{-1}$)$^2$ and $g_A$ (5.5 cm$^{-1}$)[48] are the corresponding gain coefficients, $\eta_A$ is the acceptor quantum yield (0.5),[49] $L$ is a coefficient associated to the lineshape function (=$1/\pi$ for a Gaussian profile), and $\Omega$ is the aperture angle of the emission cone. The resulting behaviour of the emission intensities is displayed in Fig. 6 as a function of the relative molar concentration of acceptors and donors, $\phi$. Considering a value of the intensity equal to ~10$^{-5}$ times the saturation intensity as threshold value for stimulated emission in the light-emitting fibers, we find that for $\phi$ =0.6, namely for a CrV content of 0.31%, the donor is still above threshold for stimulated emission,

while for $\phi$ =1.3 (CrV content=0.50%) the acceptor component is already above its own

threshold for ASE. This makes very likely the occurrence of an intermediate regime at $\phi \cong$0.6-

0.7, with tight ASE-FRET interplay and where the two components show ASE simultaneously.

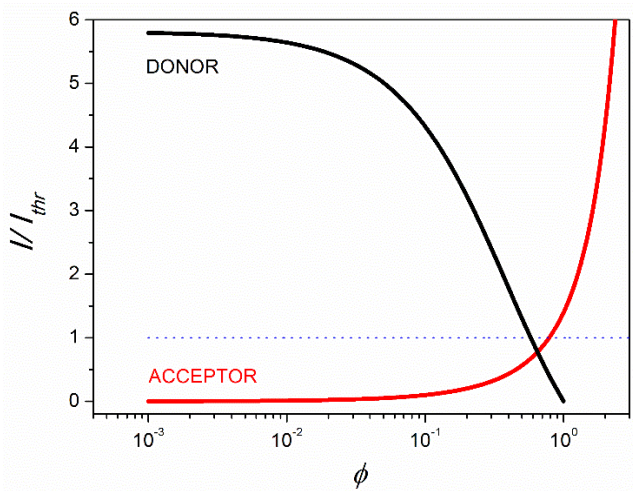

**Figure 6.** Calculated ASE intensity from the donor (black line) and from the acceptor (red line) in the light-emitting fibers, vs. acceptor/donor relative concentration. Used parameters: $R_0/a$=1.1, $\Omega$=2$\pi$×10$^{-5}$ rad (Eq. 4a and 4b). $\Omega$ is evaluated taking into account the acceptance angle of the experimental collection system, which includes an optical fiber with a core of 200 μm at 2 cm from the sample edge. The dotted horizontal line indicates the threshold intensity.

Corresponding to a measured value of CrV content ≲ 0.50%, this is in agreement with our experiments, where the influence of FRET process occurring between two dyes is visible especially for the sample containing equal amounts in weight of Rh6G and CrV, with stimulated emission from both dyes clearly observed (Fig. 5a and Fig. S5b). For these light-emitting fibers, the threshold of the CrV band is the lowest and 2.5 times lower value then for sample containing only the CrV dye, under the same excitation conditions. Without FRET-assisted stimulated emission, the threshold level should be expected at higher level than for fibers with the CrV dye only, since it would compete with Rh6G for the same excitation photons. The impact of FRET on the stimulated emission threshold is made clear in Fig. 7a and b, where we plot the experimental thresholds vs. FRET efficiency for the CrV and Rh6G dyes, respectively. Indeed, the threshold for the stimulated emission of CrV (acceptor) is being significantly decreased when the FRET efficiency reaches the highest value (Fig. 7a), while the threshold value for the donor (Rh6G) concomitantly increases (Fig. 7b).

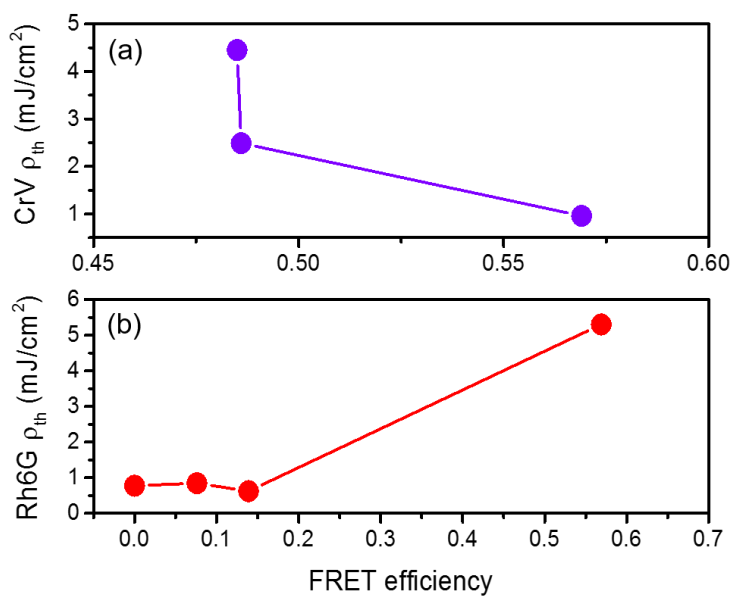

**Figure 7.** Threshold of stimulated emission vs. FRET efficiency for CrV-acceptor (a) and Rh6G-donor (b). The solid lines are guides to eye.

    *3.4. CBS in the disordered electrospun fibers.* Electrospun non-wovens are quite optically dense and non-transparent, with a relevant role in optical properties also played by light scattering. CBS experiments help in elucidating if incoherent random lasing is also taking place in the electrospun material. An exemplary cone of light backscattered from an electrospun non-woven is presented in Fig. 8a. The transport mean free path, $L_t$, for photons in the complex molecular material was calculated according to Eq. (2) and found to weakly vary around (19.1 ±

0.9) µm (Fig. 8). This is of the order of the average voids among fibers, namely pores delimited by fibers and roughly visualized in two dimension by SEM micrographs as that in Fig. 2e, and comparable to those found for other disordered materials used for incoherent random lasing, such as latex nanoparticles suspended in Rh6G solutions or grinded solid-state laser materials.[50] With a used excitation spot having a diameter of 3 mm ($>>L_t$), one concludes that incoherent random lasing might occur also from light-emitting fibers. The quite large area of excitation supports a great number of modes that overlap themselves resulting in blurred emission spectra as typical of ASE and incoherent random lasing. Additionally fast randomization of photons phase after travelling over $L_t$ also supports the incoherent feedback. These effects, based on the diffusion of light generated by electrospun fibers, might be beneficial in view of developing various lightening applications, as was recently shown for increasing the efficiency of organic light-emitting devices.[51] By exploiting FRET-assisted lasing, it would also be possible to achieve very bright light sources with stable emission performance.

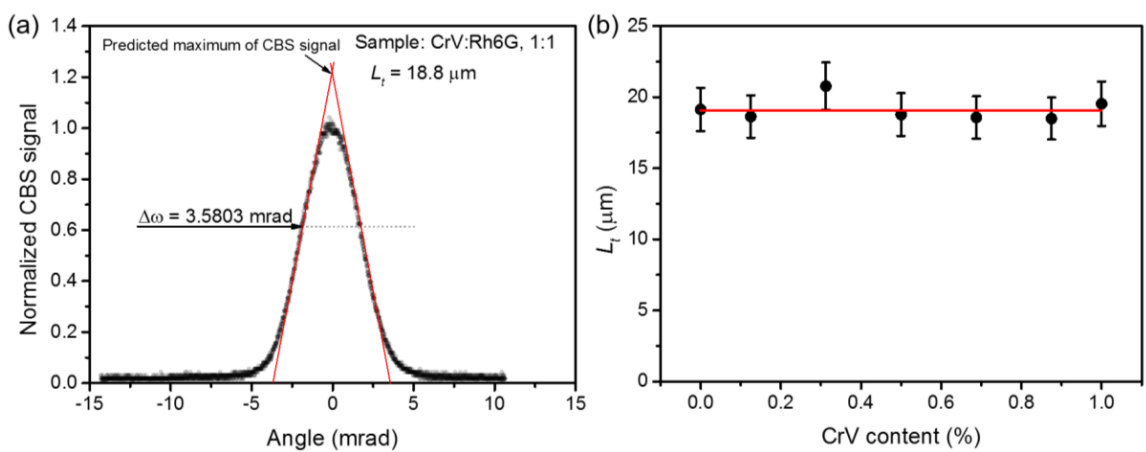

**Figure 8.** (a) Exemplary CBS signal for fibers with 0.5% Rh6G and 0.5% CrV. (b) Mean transport free path, $L_t$, measured for samples with different content of CrV.

## 5. Conclusions

To conclude we have shown that electrospun polymer fibers doped with FRET-coupled laser systems can be efficient ASE sources, with incoherent random lasing features. The use of effective FRET pairs (Rh6G and CrV) allows the threshold conditions of stimulated emission to be engineered, as well as the resulting spectral shape to be tailored through the FRET-stimulated emission interplay. The tunability of the optical gain from electrospun fibers can be significantly extended in this way, achieving stimulated emission over a range exceeding 100 nm. The transport mean free path, $L_t$, for photons is of $(19.1 \pm 0.9)$ μm, matching the inter-fiber separation distance obtained by electrospinning, which is another property in principle tailorable through the process. Approaches to utilize light-emitting and lasing non-wovens operating in incoherent regime include applications in lightening, sensing, and display technologies as efficient, bright and stable active macromolecular materials.

**Author Information**

**Corresponding Author**

Dario Pisignano. E-mail address: dario.pisignano@unipi.it

**Associated Content**

**Supporting Information.**

Emission and excitation spectra, further data on ASE intensity depencence on pumping fluence, set-up schematics and parameters of performed fits are available free of charge via the Internet at http://pubs.acs.org.

**Acknowledgements**

The authors would like to thank the Polish National Science Center for financial support (Grant no. 2013/09/D/ST4/03780, 2013/11/N/ST4/01488, and 2016/21/B/ST8/00468) and Wroclaw University of Technology for sharing facilities and equipment. The research leading to these results has received funding from the European Research Council under the European Union's Seventh Framework Programme (FP/2007-2013)/ERC Grant Agreements n. 306357 (ERC Starting Grant "NANO-JETS").

# Interplay of Stimulated Emission and Fluorescence Resonance Energy Transfer in Electrospun Light-Emitting Fibers

*Lech Sznitko,[†] Luigi Romano,[§,††] Andrea Camposeo,[††] Dominika Wawrzynczyk,[†] Konrad Cyprych,[†] Jaroslaw Mysliwiec,[†] and Dario Pisignano[§,††,¶,*]*

[†] Wroclaw University of Science and Technology, Wybrzeze Wyspianskiego 27, 50-370 Wroclaw, Poland

[§] Dipartimento di Matematica e Fisica "Ennio De Giorgi", Università del Salento, via Arnesano I-73100, Lecce, Italy

[††] NEST, Istituto Nanoscienze-CNR, Piazza S. Silvestro 12, I-56127 Pisa, Italy

[¶] Dipartimento di Fisica, Università di Pisa and CNR Istituto Nanoscienze, Largo B. Pontecorvo 3, I-56127 Pisa, Italy

[*] Corresponding author. E-mail: dario.pisignano@unipi.it

SUPPORTING INFORMATION

**Table of Content**

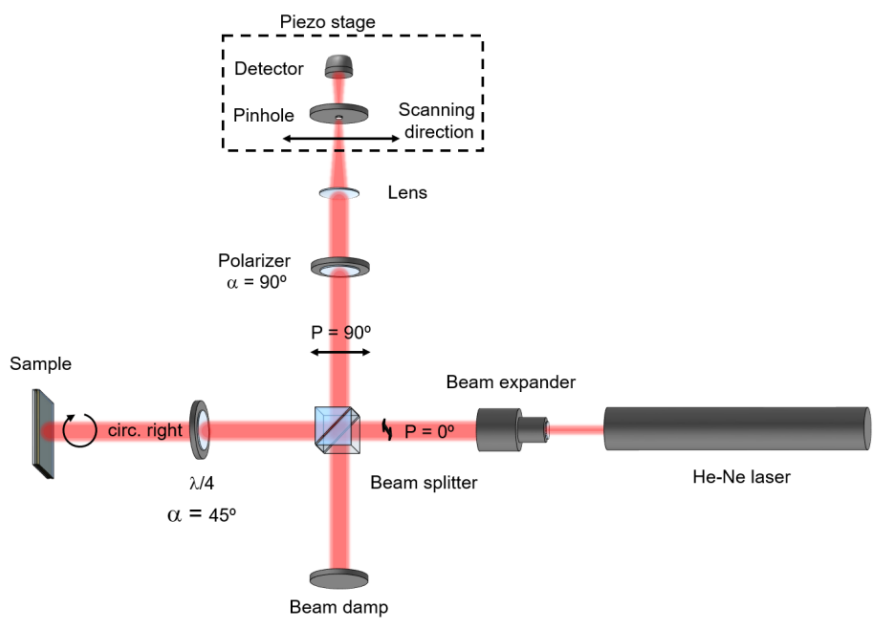

**Figure S1.** Experimental set-up for CBS measurements.

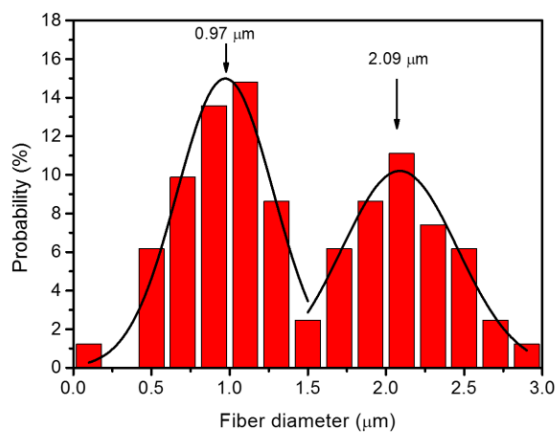

**Figure S2.** Histogram of fiber sizes, fitted with Gaussian distributions. Data are obtained from SEM micrographs of samples with 0.5% Rh6G. Two size distributions are clearly visible in histogram, corresponding to the ribbon shape of obtained fibers.

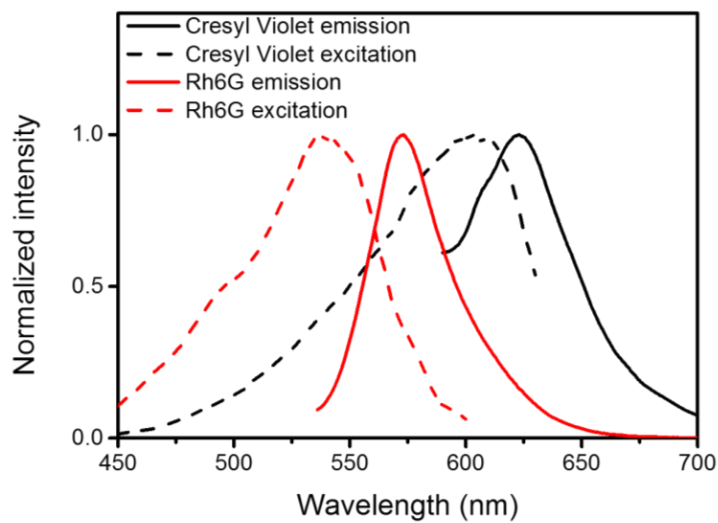

**Figure S3.** Normalized excitation and emission spectra of CrV (black lines) and Rh6G (red lines) embedded in PMMA fibers.

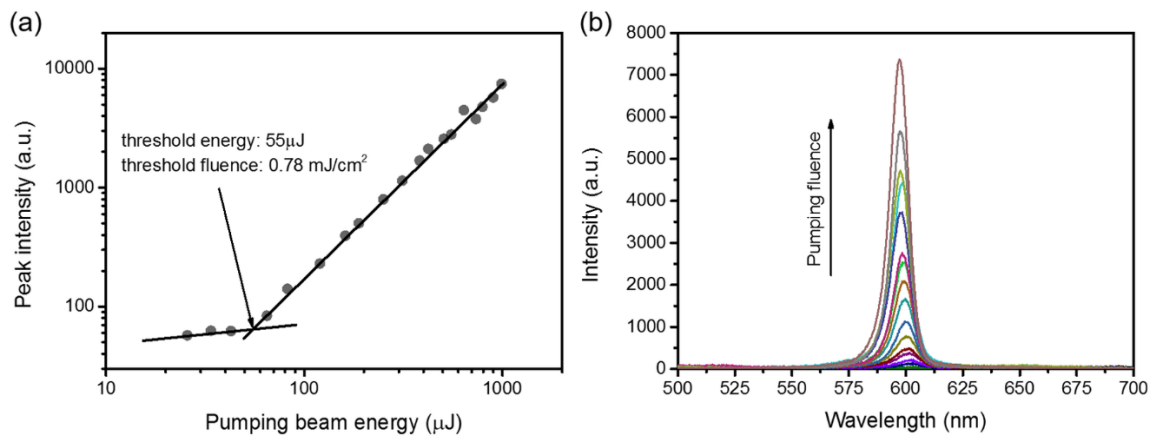

**Figure S4.** Emission peak intensity versus pumping energy (a) and stimulated emission spectra (b) obtained for sample *S*1.

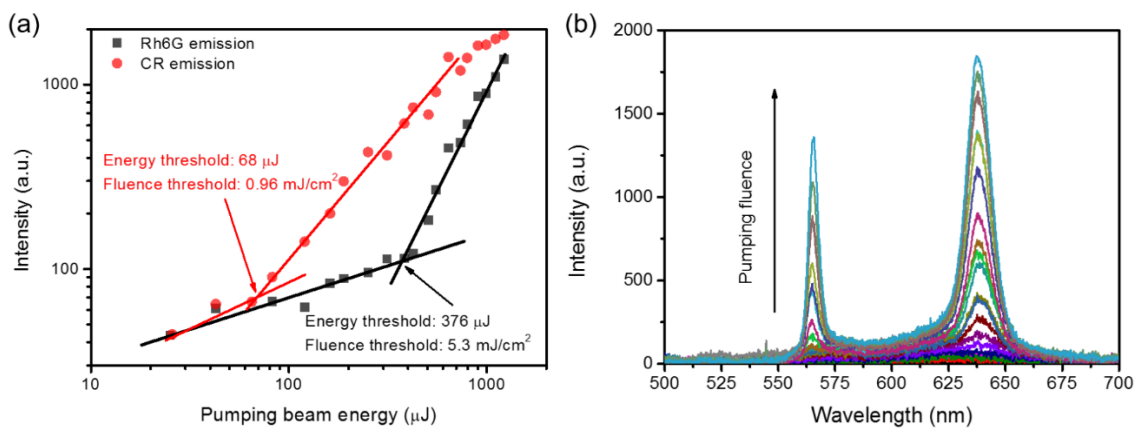

**Figure S5.** Emission peak intensity versus pumping energy (a) and stimulated emission spectra (b) obtained for sample S4. Clearly visible two stimulated emission bands.

**Table S1.** Bi-exponential decay function parameters obtained by fitting the luminescence lifetime decays.

| CrV content (%) | Rh6G content (%) | $t_1$ (ps) | $A_1$ (Arb. Un.) | $t_2$ (ps) | $A_2$ (Arb. Un.) | $\eta$ (%) |
|---|---|---|---|---|---|---|
| 0.00 | 1.00 | 1934 | 2717 | 14 | 86 | 0 |
| 0.13 | 0.87 | 1837 | 3018 | 61 | 39 | 7.5 |
| 0.31 | 0.69 | 1558 | 2726 | 53 | 47 | 13.9 |
| 0.50 | 0.50 | 566 | 1531 | 65 | 35 | 57.0 |
| 0.69 | 0.31 | 690 | 1806 | 64 | 36 | 48.5 |
| 0.87 | 0.13 | 712 | 1941 | 71 | 29 | 48.6 |